\documentclass[12pt]{article}
\usepackage{epsfig}

\def\nn{\nonumber}
\def\im{\mbox{ Im }}

\def\eg{{\em e.g.}}

\begin{document}

\title{
NUCLEON SUM RULES IN SYMMETRIC AND ASYMMETRIC NUCLEAR MATTER
\thanks{
This work is supported by the RFBR - 06-02-16353}}

\author{\underline{V.~A.~SADOVNIKOVA},  E.~G.~DRUKAREV\\
AND M.~G.~RYSKIN \\
St. Petersburg Nuclear Physics Institute \\
Gatchina 188300, Russia\\
E-mail: sadovnik@thd.pnpi.spb.ru}

\maketitle

\begin{abstract}
We calculate the nucleon parameters in  isospin symmetric and
asymmetric nuclear matter using the QCD sum rules.  The higher
moments of the nucleon structure functions are included. The
complete set of the nucleon expectation values of the four-quark
operators is employed.  We analyze the role of the lowest order
radiative corrections beyond the logarithmic approximation.
\end{abstract}

\markboth{\large \sl \underline{V.~A.~SADOVNIKOVA}
\& E.~G.~DRUKAREV
\& M.~G.~RYSKIN
\hspace*{2cm} HSQCD 2005} {\large \sl \hspace*{1cm}
HSQCD 2005 PROCEEDINGS}

\section{Introduction}
We investigate the vector and scalar self-energies of nucleons in
nuclear matter composed by the neutrons and protons, distributed
with densities $\rho_n$ and $\rho_p$. We calculate the dependence
on the total density $\rho\ =\ \rho_p+\rho_n$ and on the asymmetry
parameter $\beta\ =\ (\rho_n-\rho_p)/(\rho_p+\rho_n)$.

The QCD sum rules were invented in paper \cite{1} to express the
hadron parameters through the vacuum expectation values of QCD
operators. Being initially used for the mesons, the method was
expanded in \cite{2} to the description of the baryons. The approach
succeeded in describing the static characteristics as well as some
of the dynamical characteristics of the hadrons in vacuum --- see,
\eg\, the reviews \cite{4}.

Later the QCD sum rules were applied for investigation of modified
nucleon parameters in the symmetric nuclear matter \cite{7}. They
were based on the Borel-transformed dispersion relation for the
function $\Pi_m(q)$ describing the propagation of the system with
the quantum numbers of the nucleon (the proton) in the nuclear
matter. Considering nuclear matter as a system of $A$ nucleons
with momenta $p_i$, one introduces the vector $p
 =(\Sigma p_i)/A $,
which is thus $p\approx(m,0)$ in the rest frame of the matter. The
function $\Pi_m(q)$ can be presented as
$\Pi_m(q)=\Pi_m(q^2,\varphi(p,q))$ with the arbitrary function
$\varphi(p,q)$ being kept constant in the dispersion relations in
$q^2$.

The general form of the function $\Pi_m$ can thus be presented as
\begin{equation} \label{7}
\Pi_m(q)\ =\ q_\mu\gamma^\mu\Pi^q_m(q^2,s)+p_\mu\gamma^\mu\Pi^p_m
(q^2,s)+I\Pi^I_m(q^2,s)\ .
\end{equation}
The in-medium QCD sum rules are the Borel-transformed dispersion
relations for the components $\Pi^j_m(q^2,s)$ $(j=q,p,I)$
\begin{equation} \label{8}
\Pi^j_m(q^2,s)\ =\ \frac1\pi\int\frac{\im\Pi^j_m(k^2,s)}{k^2-q^2}\
dk^2\ .
\end{equation}
The spectrum of the function $\Pi_m(q)$ is much more complicated
than that of the function $\Pi_0(q)$ describing the propagation of
the system with the quantum numbers of the nucleon in the vacuum.
The choice of the function $\varphi(p,q)$ is dictated by the
attempts to separate the singularities connected with the nucleon in
the matter from those connected with the properties of the matter
itself. Since the latter manifest themselves as singularities in the
variable $s=(p+q)^2$, the separation can be done by putting
$\varphi(p,q)=(p+q)^2$ and by fixing \cite{8} $ \varphi(p,q)\ =\
(p+q)^2\ \equiv\ s\ =\ 4m^2$ ($m$ is the nucleon mass).

By using Eq.~(\ref{8}) the characteristics of the nucleon in
nuclear matter can be expressed through the in-medium values of
QCD condensate. The possibility of extension of  "pole +
continuum" model \cite{1,2} to the case of finite densities was
shown in \cite{8}--\cite{10}.

The lowest order of OPE of the lhs of Eq.~(\ref{8}) can be
presented in terms of the vector and scalar condensates \cite{8},
\cite{10}. Vector condensates $v^i_\mu=\langle M|\bar
q\,^i\gamma_\mu q^i|M\rangle$ of the quarks with the flavor $i$
($|M\rangle$ denotes the ground state of the matter) are the
linear functions of the nucleon densities $\rho_n$ and $\rho_p$.
In the asymmetric matter both SU(2) symmetric and asymmetric
condensates
$$
v_\mu=\langle M|\bar u(0)\gamma_\mu
u(0)+\bar d(0)\gamma_\mu d(0)|M\rangle = v^u_\mu+v^d_\mu
$$
$$
v^{(-)}_\mu
=\langle M|\bar u(0)\gamma_\mu u(0)-\bar d(0)\gamma_\mu
d(0)|M\rangle=v^u_\mu-v^d_\mu
$$
obtain nonzero values.  In the rest frame of the matter $v^i_\mu
=v^i\delta_{\mu0}$, $v_\mu=v\delta_{\mu0}$,
$v^{(-)}_\mu=v^{(-)}\delta_{\mu0}$. We can present $v^i\ =\
\langle n|\bar q^i\gamma_0 q^i|n\rangle\rho_n +\langle p|\bar
q^i\gamma_0 q^i|p\rangle\rho_p.$ The values $\langle N|\bar
q^i\gamma_0 q^i|N\rangle$ are just the numbers of the valence
quarks in the nucleons $\langle n|\bar u\gamma_0 u|n\rangle=
\langle p|\bar d\gamma_0 d|p\rangle=1$, $\langle p|\bar u\gamma_0
u|p\rangle= \langle n|\bar d\gamma_0 d|n\rangle=2$, and thus
\begin{equation}
\label{10}
v^u=\ \rho_n+2\rho_p=\ \rho\left(\frac32-\frac\beta2\right), \quad
v^d=\ 2\rho_n+\rho_p=\ \rho\left(\frac32+\frac\beta2\right).
\end{equation}
Hence, we obtain $v(\rho)=v_N\rho, v^{(-)}(\rho,\beta) = \beta
v^{(-)}_N\rho$ with $v_N = 3$, $v^{(-)}_N=-1$.

The lhs of Eq.~(\ref{8}) contains the SU(2) symmetric scalar
condensate $\kappa_m(\rho)\ =\ \langle M|\bar u(0)u(0)+\bar
d(0)d(0)|M\rangle$ , and SU(2) asymmetric one
$\zeta_m(\rho,\beta)\ =\ \langle M|\bar u(0)u(0)-\bar
d(0)d(0)|M\rangle$. These condensates can be presented as
\begin{equation}\label{15}
 \kappa_m(\rho)=\kappa_0+\kappa(\rho)\ ,\quad
 \kappa(\rho)=\kappa_N\rho+\ldots\ , \quad \kappa_N=\langle N|\bar
uu+\bar dd|N\rangle\ ,
\end{equation}
 $\kappa_0=\kappa_m(0)$ is the vacuum value, and
\begin{equation}
\zeta_m(\rho,\beta)\ =\ -\beta(\zeta_N\rho+\ldots)\ , \quad \zeta_N=\langle
p|\bar uu-\bar dd|p\rangle\ .
\label{16}
\end{equation}
The dots in the rhs of Eqs.~(\ref{15}) and (\ref{16}) denote the
terms, which are nonlinear in $\rho$. In the gas approximation
such terms should be omitted. The SU(2) invariance of vacuum was
assumed in Eq.~(\ref{16}). The expectation value $\kappa_N$ is
related to the $\pi N$ sigma term $\sigma_{\pi N}$ \cite{11}.

The gluon condensate
$$
 g_m(\rho)\ =\ \langle M|\frac{\alpha_s}\pi\ G^2(0)|M\rangle\ =\
g_0+g(\rho)\ =    g_0\ +\ g_N\rho+\ \ldots
$$
$g_0 = g_m(0)$ is the vacuum value and $g_N\ =-\frac89\,m\ $
obtained in a model-independent way.

We shall analyze the sum rules in the gas approximation. It is a
reasonable starting point, since the nonlinear contributions to the
most important scalar condensate $\kappa(\rho)$ are relatively small at
the densities of the order of the phenomenological saturation density
$\rho_0=0.17\,\rm fm^{-3}$  of the symmetric matter \cite{10}.

In the second part of the talk we discuss the role of the
radiative corrections.
 The analysis  [2] included also the most important radiative corrections,
 in which the coupling constant $\alpha_s$ is enhanced by the ``large
logarithm" $\ln q^2$. The corrections $(\alpha_s\ln q^2)^n$ have
been included in to all orders for the leading OPE terms. This
approach provided us with good results for the nucleon mass  and
for the other characteristics of nucleons.

However, inclusion of the lowest order radiative corrections
beyond the logarithmic approximation made the situation somewhat
more complicated. A numerically large coefficient of the lowest
radiative correction to the leading OPE of the polarization
operator $\Pi_0(q)$ was obtained in \cite{CJ}. A more consistent
calculation \cite{CJ} provided this coefficient to be about 6.
Thus, the radiative correction increases this term by about 50\%
at $|q^2|\sim1\rm\,GeV^2$, which are actual for the SR analysis.
This uncomfortably large correction is often claimed as the most
weak point of the SR approach \cite{CJ}.

The radiative corrections of the order $\alpha_s\ln q^2$ and
$\alpha_s$ for the contributions up to $q^{-2}$ have been
calculated in \cite{OP}.

The further development of the nucleon SR in nuclear matter needs
the calculation of the radiative corrections. This work is in
progress and now we have present  the analysis of the role of the
radiative corrections in vacuum \cite{AS}.

\section{Sum rules in nuclear matter}

We present the nucleon propagator in nuclear matter as
\begin{equation}
G^{-1}_N\ =\ q_\mu\gamma^\mu-m - \Sigma \,,  \quad \Sigma\ =\
q_\mu\gamma^\mu\Sigma_q+p_\mu\gamma^\mu\,\frac{\Sigma_p}m+
\Sigma_s\  \label{21}
\end{equation}
with the total self-energy $\Sigma$ . We shall use the QCD sum
rules for the calculation of the nucleon characteristics
\begin{equation}
\Sigma_v=\frac{\Sigma_p}{1-\Sigma_q}\ , \quad
m^*=\frac{m+\Sigma_s}{1-\Sigma_q}\ , \quad \Sigma^*_s=m^*-m\ ,
\label{23}
\end{equation}
identified with the vector self-energy,  Dirac effective mass
and the effective scalar self-energy --- see \eg\ \cite{22}.
\begin{equation}
G_N\ =\ Z_N\cdot\frac{q_\mu\gamma^\mu-p_\mu\gamma^\mu
(\Sigma_v/m)+m^*}{q^2-m^2_m}
\label{48}
\end{equation}
with $\Sigma_v$ and $m^*$ defined by Eq.~(\ref{23}). The new
position of the nucleon pole is
\begin{equation}
m^2_m\ =\
\frac{(s-m^2)\Sigma_v/m-\,\Sigma^2_v+m^{*2}}{1+\Sigma_v/m}\ ,
\quad Z_N\ =\ \frac1{(1-\Sigma_q)(1+\Sigma_v/m)}\ . \label{49}
\end{equation}
We present also the result for the single-particle potential
energies
\begin{equation}
U\ =\ \Sigma^*_s+\Sigma_v\ . \label{24}
\end{equation}
We trace the dependence of these characteristics on the total
density $\rho$ and on the asymmetry parameter.

The Borel-transformed sum rules take the form
\begin{eqnarray}
\label{52}
L^q_m(M^2,W^2_m) &=& \Lambda_m(M^2)\ ,\\
\label{53}
L^p_m(M^2,W^2_m) &=& -\Sigma_v\Lambda_m(M^2)\ ,\\
\label{54}
L^I_m(M^2,W^2_m) &=& m^*\Lambda_m(M^2)
\end{eqnarray}
with $ \Lambda_m(M^2)\ =\ \lambda^{*2}_m e^{-m^2_m/M^2}$, where
$\lambda^{*2}_m$ is the effective value of the nucleon residue in
nuclear matter, $M$ is Borel mass ($0.8\leq M^2\leq1.4$GeV$^2$).

We shall include subsequently the contributions of three types. The
terms $\ell^j_m(M^2)$ $(j=p,q,I)$ stand for the lowest order local
condensates. These contributions are similar to simple exchanges by
isovector vector and scalar mesons between the nucleon
 and the nucleons of the matter. The terms
$u^j_m(M^2)$ are caused by the nonlocalities of the vector
condensate. They correspond to the account of the form factors in
the vertices between the isovector mesons couple to the nucleons.
Finally, $\omega^j(M^2)$ describes the contributions of the
four-quark condensates. They correspond to the two-meson exchanges
(or to exchanges by four-quark mesons, if there are any) and to
somewhat more complicated structure of the meson-nucleon vertices.
Thus we present the lhs of Eqs. (\ref{52})--(\ref{54}) as
\begin{equation} \label{60}
L^j_m\ =\ \ell^j_m+u^j_m+w^j_m .
\end{equation}
Actually, we shall solve the sum rules equations, subtracting the
vacuum effects \cite{anm} :
\begin{equation} \label{61}
L^j\ =\ \ell^j+u^j+w^j
\end{equation}
with $\ell^j=\ell^j_m-\ell^j_0$, $u^j=u^j_m$,
$\omega^j=\omega^j_m-\omega^j_0$, while $\ell^j_0$ and $\omega^j_0$ are
the corresponding contributions in the vacuum case.

\vspace{0.5cm}

\noindent
{\bf A. Local condensates of the lowest dimensions.}\\
The terms $\ell^i$ have the form:
\begin{eqnarray}
 && \ell^q\ =\
f^q_v(M^2,W^2_m)v^q(\rho)+f^q_g(M^2,W^2_m)g(\rho),
\\
&& \ell^p\ =\ f^p_v(M^2,W^2_m)v^p(\rho,\beta)\ ,
\\
&& \ell^I\ =\ f^I_\kappa (M^2,W^2_m)t^I(\rho,\beta)\ ,
\end{eqnarray}
with the dependence on $\rho$ and $\beta$ being contained in the
factors
\begin{equation}
v^q(\rho)=3\rho, \quad
v^p(\rho,\beta)=3\rho\left(1-\frac\beta4\right), \quad
t^I(\rho,\beta)=\rho(\kappa_N+\zeta_N\beta),
\end{equation}
and the function $g(\rho)$, given by Eqs.~(\ref{15}) and
(\ref{16}). The other functions are \cite{16}
\begin{eqnarray}
&& f^q_v(M^2,W^2_m)
=-\frac{8\pi^2}3\frac{(s-m^2)M^2E_{0m}-M^4E_{1m}}{m\,L^{4/9}},
\nn \\
&& f^q_g(M^2,W^2_m)=\frac{\pi^2 M^2E_{0m}}{L^{4/9}},
\nn \\
&& f^p_v(M^2,W^2_m)=-\frac{8\pi^2}3\ \frac{4M^4\,E_{1m}}{L^{4/9}}\
,
\nn\\
&& f^I_\kappa(M^2,W^2_m)=-4\pi^2M^4E_{1m}\ . \label{72}
\end{eqnarray}
The notation $E_{km}\,(k=0,1)$ means that the functions
$E(x)=1-e^{-x}\Sigma_{k=0}^{n}\frac{x^n}{n}$ depend on the ratio
$W^2_m/M^2$. The factor $L=\ln(M^2/\Lambda^2)
/\ln(\nu^2/\Lambda^2)$, where $\Lambda=0.15$~GeV, $\nu= 0.5$~GeV.

\vspace{0.5cm}

\noindent
{\bf B. Inclusion of the nonlocal condensates.}\\
Finally, the higher moments and higher twists of the nucleon
structure functions provide the contributions $u^i$ to the $L^i$
of the sum rules --- Eq.~(\ref{61})
\begin{eqnarray}
&&
u^q(M^2)\ =\ (u^q_{N,1}(M^2) + \beta u^q_{N,2}(M^2))\rho\ ,
\nn \\
&& \quad
u^q_{N,1}(M^2)=\frac{8\pi^2}{3L^{4/9}m}\left[
-\frac52 m^2 M^2 E_{0m} \langle\eta\alpha\rangle
+\frac32 m^2 (s-m^2)\langle\xi\rangle\right]\ ,
\nn\\
&&\quad  u^q_{N,2}(M^2)=\frac{8\pi^2}{3L^{4/9}m}\left[
\frac32\,m^2M^2E_{0m}(\langle\eta^u\alpha\rangle
-\langle\eta^d\alpha\rangle) \right] ;
\label{85}
\\
\nn\\
\nn\\
&&  u^p(M^2)\ =\ (u^p_{N,1}(M^2) + \beta u^p_{N,2}(M^2))\rho\ ,
\nn\\
&& \quad
u^p_{N,1}(M^2)=\frac{8\pi^2}{3L^{4/9}}\bigg[-5\left(M^4 E_{1m}
-(s-m^2)M^2 E_{0m}\right)\langle\eta\alpha\rangle
\nn\\
&&
\quad -\ \frac{12}5 m^2M^2E_{0m}\langle\eta\alpha^2\rangle
+\ \frac{18}5m^2M^2E_{0m}\langle\xi\rangle\bigg]\ ,
\nn \\
&& \quad
u^p_{N,2}(M^2)=\frac{8\pi^2}{3L^{4/9}}\bigg[
 3\left(M^4E_{1m}-(s-m^2)M^2E_{0m}\right)
\langle(\eta^u-\eta^d)\alpha\rangle
\nn\\
\label{86m}
&&
\quad+\ \frac95\,m^2M^2E_{0m}\langle(\eta^u-\eta^d)\alpha^2\rangle
-\frac{27}{10}\,m^2M^2E_{0m}
\langle(\xi^u-\xi^d)\rangle \bigg] ;
\\
\nn\\
\nn\\
&&
u^I(M^2)\ =\ 0\ .
\label{85a}
\end{eqnarray}
Here we denote $\xi^u = -0.24$, $\xi^d = -0.09$ \cite{25} and $\xi
= \xi^u + \xi^d$.  We use the structure functions $\eta$, obtained
in \cite{30} for the calculation of the terms $u^q$ and $u^p$.
Here we denote $\langle f\rangle=\int^1_0 d\alpha f(\alpha)$.

\vspace{0.5cm}

\noindent
{\bf C. Inclusion of the four-quark condensates.}\\
The calculations of the contributions of the four-quark condensates
require the model assumption on the structure of the nucleon. The
complete set of the four-quark condensate was obtained in \cite{17}
by using the perturbative chiral quark model (PCQM). There are three
types of contributions to the four-quark condensate in the framework
of this approach.  All four operators can act on the valence quarks.
Also, four operators can act on the pion. There is  a possibility
that two of the operators act on the valence quarks while the other
two act on the pions. Using the complete set of the nucleon
four-quark expectation values \cite{17}, we obtain
\begin{equation}\label{104}
(\Pi)_{4q}\ =\ \left(A^q_{4q}(\beta)\frac{\hat q}{q^2}+A^p_{4q}(\beta)
\frac{(pq)}{m^2}\frac{\hat p}{q^2}+A^I_{4q}(\beta)m\frac I{q^2}
\right)\frac a{(2\pi)^2}\rho
\end{equation}
with $a\ =\ -(2\pi)^2\langle 0|\bar uu|0\rangle$. The calculations
give
\begin{equation}
\label{106}
A^q_{4q}=-0.11-0.21\beta, \quad A^p_{4q}=-0.57+0.09\beta, \quad
A^I_{4q}=1.90-0.92\beta\ .
\end{equation}
The contributions of the four-quark condensates to the lhs of the
Borel transformed sum rules (\ref{61})  can be presented  as
\begin{eqnarray} &&
\omega^j=\omega^j_N\rho\ , \quad \omega^j_N=A^j_{4q}(\beta)f^j_{4q}\ ,
\nn \\
&&  f^q_{4q}=-8\pi^2 a , \quad f^p_{4q}=-8\pi^2\frac{s-m^2}{2m} a ,
\quad f^I_{4q}=-8\pi^2 m a.
\label{108}
\end{eqnarray}

In Fig.~1 we show results for effective nucleon mass $m^*$
(\ref{23}) and for the single-particle potential energies (\ref{24})
in symmetric and asymmetric nuclear matter. During calculation we
use the value $\kappa_N=8$ (\ref{15}), and $\zeta_N=0.54$ (\ref{16})
in Eq.(19)  \cite{anm}.

\begin{figure}
\centering{\epsfig{figure=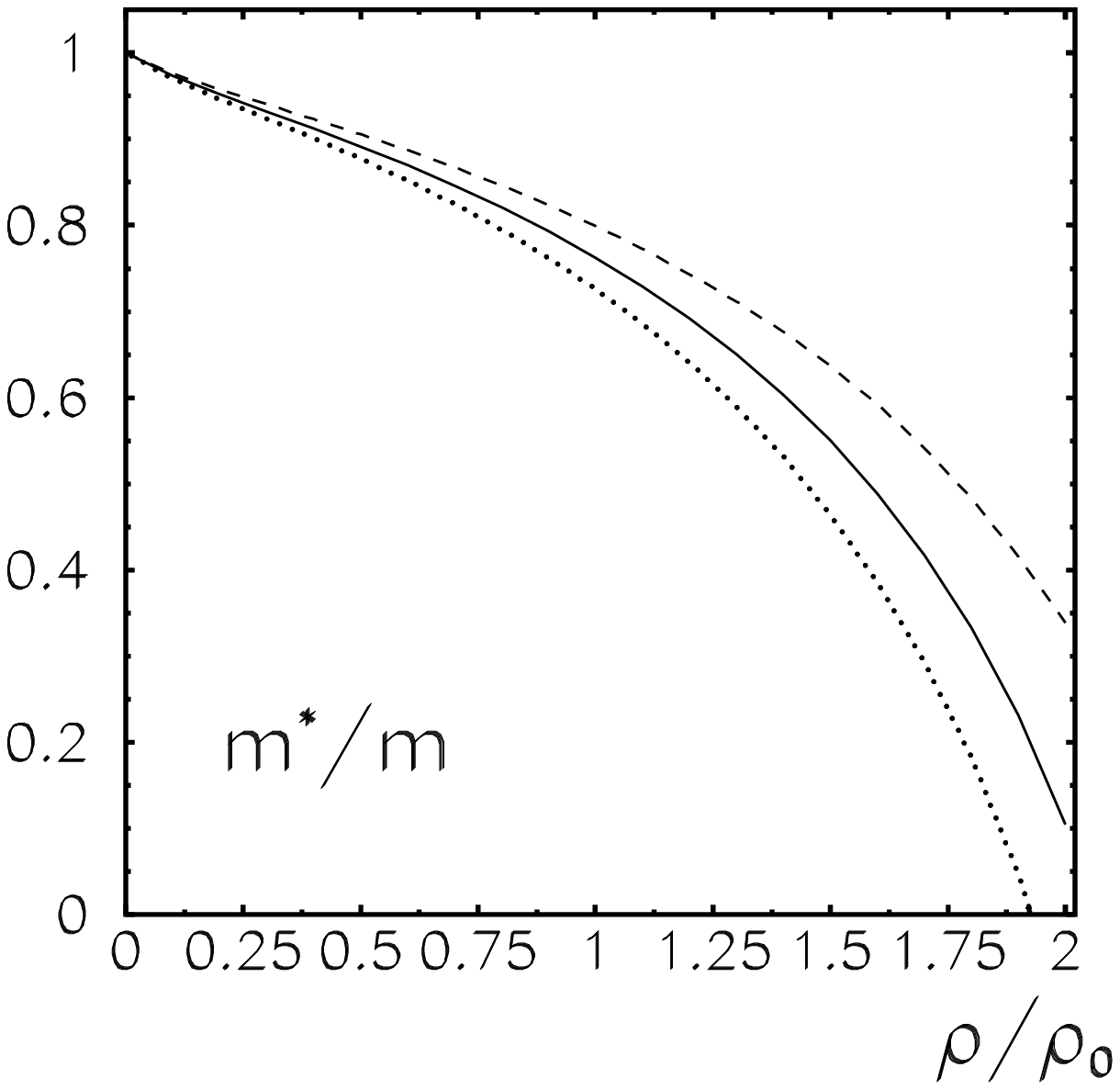,width=5cm}
\epsfig{figure=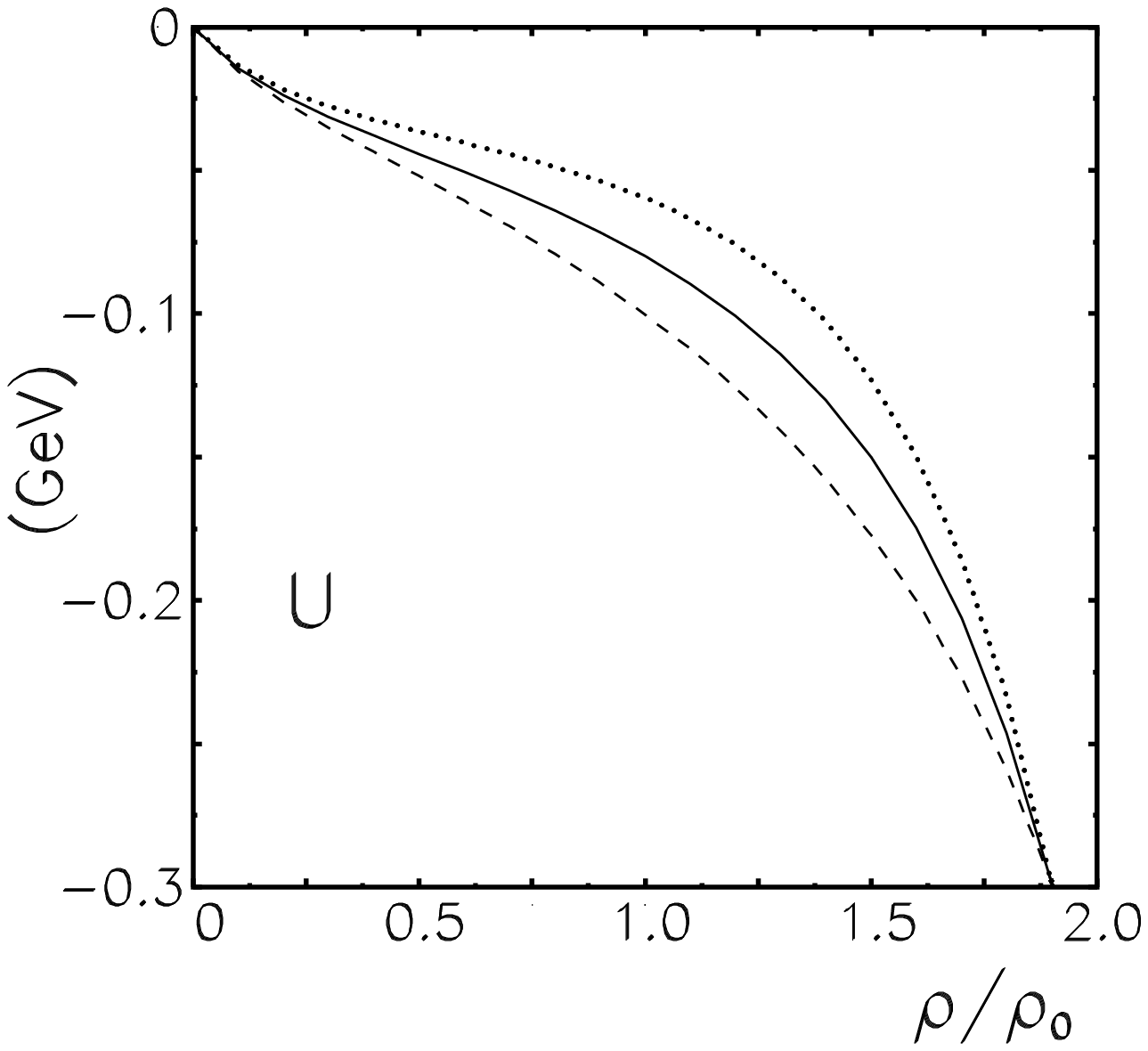,width=5.5cm}}%
 \caption{
 Results for $m^*$ and $U(\rho)$ both in symmetric and neutron
matter. Solid curves are for symmetric matter; proton (dashed) and
neutron (dotted) curves are  for the neutron matter. }
\end{figure}

\section{The role of radiative corrections}
We have made \cite{AS} the analysis of  radiative corrections to
the nucleon SR in vacuum. The lowest OPE terms of the operators
$\Pi^q_0(q^2)$ and $\Pi^I_0(q^2)$ in vacuum  can be presented as
\begin{equation}
\Pi^q_0 = A_0+A_4+A_6+A_8; \quad \Pi^I_0=B_3+B_7+B_9\ .
\end{equation}
Here the lower indices show the dimensions of the condensates,
contained in the corresponding terms, $A_0$ is the contribution of
the free quark loop.

We consider the radiative corrections to the terms $A_0$, $A_6$,
and $B_3$. The radiative corrections to the other terms are not
included since the values of the corresponding condensates are
known with poor accuracy. We find, following \cite{SH}, for the
corresponding contributions from $\alpha_s$-corrections to the
Borel transformed SR:
\begin{eqnarray}
&& \hspace*{-0.7cm} \tilde
A_0(M^2,W^2)=M^6E_2\!\left[\!1+\frac{\alpha_s}\pi\!\left(
\!\frac{53}{12}-\ln\frac{W^2}{\nu^2}\!\right)\!\right]
\nonumber\\
&& -\ \frac{\alpha_s}{\pi}\bigg[M^4W^2\!\left(1+\frac{3W^2}{4M^2}
\right)e^{-W^2/M^2}
   \nonumber\\
&&+\ M^6 {\cal E}\!\left(-{\rm W^2/ M^2}\right)\bigg],
\nonumber\\
&& \hspace*{-0.7cm} \tilde A_6(M^2,W^2)\ =\ \frac43\,a^2
\\
&&\hspace*{-0.3cm} \times
\left[\!1\!-\frac{\alpha_s}\pi\!\left(\!\frac56+\frac13\!
\left(\!\ln\frac{W^2}{\nu^2}\!+{\cal E}(-{\rm
W^2/M^2})\!\right)\!\right)\!\right]\!,
\nonumber\\
&& \hspace*{-0.7cm} \tilde B_3(M^2,W^2)\ =\
2aM^4E_1\left(1+\frac32\,\frac{\alpha_s}\pi\right) \nonumber
\end{eqnarray}
with $ {\cal E}(x) = \sum_{n=1}{x^n}/{(n\cdot n!)}$. Some terms in
(28) differ from those in \cite{SH}. The numerical difference is,
however, not very important.

The parts of equations proportional to $\alpha_s$ are the
$\alpha_s$-corrections to main terms. Now we compare the nucleon
parameters obtained as solutions of nucleon SR
 without $\alpha_s$-corrections,
$$
 m=0.930\mbox{ GeV}, \lambda^2=1.79\mbox{ GeV}^6,
 W^2 = 2.00\mbox{ GeV}^2,
$$
and with $\alpha_s$-corrections,
$$
 m=0.94\mbox{ GeV },  \lambda^2=2.00\mbox{ GeV}^6,
  W^2 = 1.90\mbox{ GeV}^2.
$$
These results are presented for $\alpha_s$=0.35.
 We show that in vacuum the radiative corrections modify mainly the values
of the nucleon residue, while that of the nucleon mass suffers
minor changes.


\begin{thebibliography}{99}
\bibitem{1} M.A. Shifman, A.I. Vainshtein and V.I.~Zakharov, Nucl.
Phys. B {\bf147}, 385, 448, 519 (1979).

\bibitem{2} B.L. Ioffe, Nucl. Phys. B {\bf188}, 317 (1981);
 B.L. Ioffe and A.V. Smilga, Nucl. Phys. B {\bf232}, 109
(1984).

\bibitem{4} M.A. Shifman, {\em "Vacuum structure and QCD sum rules"}
(North Holland, Amsterdam, 1992);
 L.J. Reinders, H. Rubinstein, S.~Yazaki, Phys. Rep.
{\bf127}, 1 (1985).

\bibitem{7} E.G. Drukarev and E.M. Levin, JETP Lett. {\bf48}, 338
(1988); Sov. Phys. JETP {\bf68}, 680 (1989).

\bibitem{8} E.G. Drukarev and E.M. Levin, Nucl. Phys. A {\bf511}, 679
(1990); Prog. Part. Nucl. Phys. {\bf27}, 77 (1991);
 E.G. Drukarev and M.G. Ryskin, Nucl. Phys. A {\bf578}, 333
(1994).

\bibitem{14} R.J. Furnstahl, D.K. Griegel and T.D.~Cohen, Phys. Rev. C
{\bf46}, 1507 (1992); X. Jin, T.D. Cohen, R.J. Furnstahl, and
D.K.~Griegel, Phys.  Rev. C {\bf47}, 2882 (1993).


\bibitem{10} E.G. Drukarev, M.G. Ryskin and V.A.~Sadovnikova, Prog.
Part. Nucl. Phys. {\bf47}, 73 (2001).

\bibitem{11} J. Gasser, Ann. Phys. {\bf136}, 62 (1981).

\bibitem{CJ} Y. Chung, H. Dosch, M. Kramer, and D.~Schall, Z. Phys.
    C~{\bf25}, 151 (1984);
 M. Jamin, Z. Phys. C {\bf37}, 635 (1988);
 D.~V. Leinweber, Ann. Phys. {\bf254}, 328 (1997).

\bibitem{OP} A.~A. Ovchinnikov, A.~A.~Pivovarov, and L.~R.~Surguladze,
   Int. J. Mod. Phys. A~{\bf6}, 2025 (1991).

\bibitem{AS} V.~A. Sadovnikova, E.~G. Drukarev, M.~G.~Ryskin, Phys.
Rev. D~{\bf72}, 114015 (2005).

\bibitem{22}  L.S. Celenza and  C.M. Shakin, {\em"Relativistic Nuclear
Physics"} (World Scientific, Philadelphia, 1986).

\bibitem{anm} E.G. Drukarev, M.G. Ryskin, V.A.~Sadovnikova,
 Phys. Rev. D {\bf70}, 065206 (2004).

\bibitem{16} E.G. Drukarev, M.G. Ryskin, V.A.~Sadovnikova,
 Th.~Gutsche and Amand~Faessler, Phys.  Rev. C {\bf69}, 065210
 (2004).
\bibitem{25} V.M. Braun and A.V. Kolesnichenko, Nucl. Phys. B {\bf283},
723 (1987).

\bibitem{30} A.D. Martin, R.G. Roberts and W.J.~Stirling, Phys. Lett. B
{\bf387}, 419 (1996).

\bibitem{17} E.G. Drukarev, M.G. Ryskin, V.A.~Sadovnikova,
V.E.~Lyubovitskij, Th.~Gutsche and A.~Faessler,
 Phys. Rev. D {\bf68}, 054021 (2003).

\bibitem{SH} H.~Shiomi and T.~Hatsuda, Nucl.~Phys. {\bf A594}, 294
(1995).


\end{thebibliography}
\end{document}